# New Transformative Possibilities for Ovonic Devices

Stanford R. Ovshinsky
Ovshinsky Innovation LLC
1050 East Square Lake Road
Bloomfield Hills, MI 48304 U.S.A.
sovshinsky@ovshinskyinnovation.com

**ABSTRACT**

The Ovonic Phase Change Memory is critical in the quest to meet the increasing commercial needs for new information systems. The important paper of DerChang Kau et al. [1], describing a stackable cross point phase change memory, resulting in a three dimensional device which includes the Ovonic Threshold Switch, therefore permits a significant expansion of our field.

I will propose and show that our basic science allows for other transformative changes that combine both threshold and phase change mechanisms in a single device. The mechanisms that I will describe have already been proven. My aim is to provide the understanding of these mechanisms so that new devices based on them can become important components that illustrate the great potential of our field.

I know of no better place to give this kind of talk than Italy where I have had some of my happiest memories of working together with Roberto Bez and his wonderful colleagues and collaborators.

## 1. INTRODUCTION

Many of you in the audience have heard me discuss through the years the fact that we can make with our Ovonic devices and mechanisms circuits that not only do non volatile memory but also do logic. Now, just recently there has been a great fuss by an important semiconductor company in which they feature devices known as memristors capable of doing both logic and memory [2]. Their claim is that they can implement what is called IMP logic operation ('logic by implication' or 'p implies q' type logic). The truth table for that kind of logic is shown in Figure 1:

| $q'$ ← | | $p\text{IMP}q$ |
|---|---|---|
| In | In | Out |
| $P$ | $q$ | $q'$ |
| 0 | 0 | 1 |
| 0 | 1 | 1 |
| 1 | 0 | 0 |
| 1 | 1 | 1 |

**Figure 1 – Truth Table**

What I want to show in this presentation is that the properties of both the OTS and OMS devices can be described in terms of them being memristors and further – that the same kind of logic (by material implication) can be performed by OTS and OMS devices, or rather three terminal versions of them.

## 2. OUTLINE

What I want to show here is that:

1/ The Ovonic Threshold Switch behaves like a memristor for voltages close to the threshold voltage.

2/ The cognitive mode of operation of the Ovonic Memory Switch (the Ovonic cognitive memory device) is a memristor device.

3/ The three terminal OMS device can execute IMP logic (p implies q) in a single device – the same logic that HP claims 3 of their memristor devices are required to execute.

4/ The three terminal OTS device in 'latching' mode can be used not only as a replacement for the access transistor to the OMS bit (as shown in the Intel paper), but to perform the same IMP logic. The difference, compared to the three terminal OMS-type logic, is that the 3-terminal OTS device does not store the result in a non-volatile fashion. On the other hand this logic can be extremely fast. (Faster than the OMS logic, because it does not involve a phase change.)

## 3. DISCUSSION

*1/ The Memristor*

First, what is a memristor?

The Memristor is also called the fourth basic circuit element (along with resistor, capacitor and inductance). Its existence was first hypothesized by Chua in 1971 [3]. In a circuit it implements the relationship between magnetic flux 'f' and electric charge 'q':

$$M(q) = df(q)/dq$$

$M(q)$ has the dimension of resistance and is called 'memristance' – or resistance with memory. The reason for that is that, being a function of electric charge, its value at any time is determined by the current through the device for all previous times. In other words it is a resistance with memory – it stores the information about its past in its value.   It should also be noted that if $f(q)$ is a linear function of q, $M(q)=R$ is a constant and the memristor becomes a 'normal' resistor. Therefore, the memristor should be defined as a non-linear resistor with memory.

*2/ The Ovonic Threshold Switch*

As we mentioned above, Leon Chua hypothesized about the existence of memristors as far back as 1971. It is remarkable, that in his original 1971 paper, he used the Ovonic Threshold Switch as one of the examples for a memristor device [3].

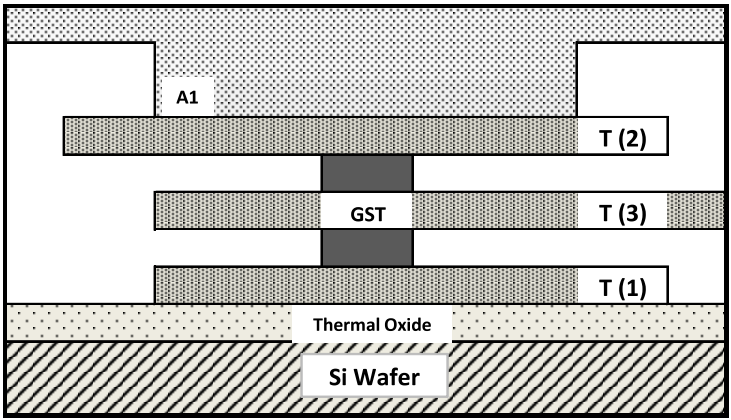
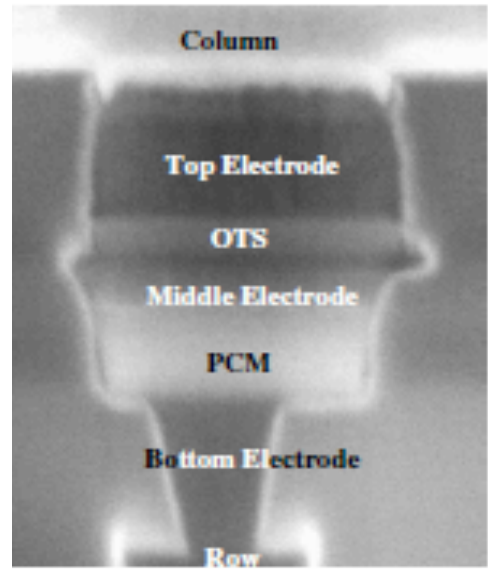

**Figure 2**
**3-terminal OTS/OMS device Memory**

**Figure 3**
**A Stackable Cross Point Phase Change**
**(from Intel)**

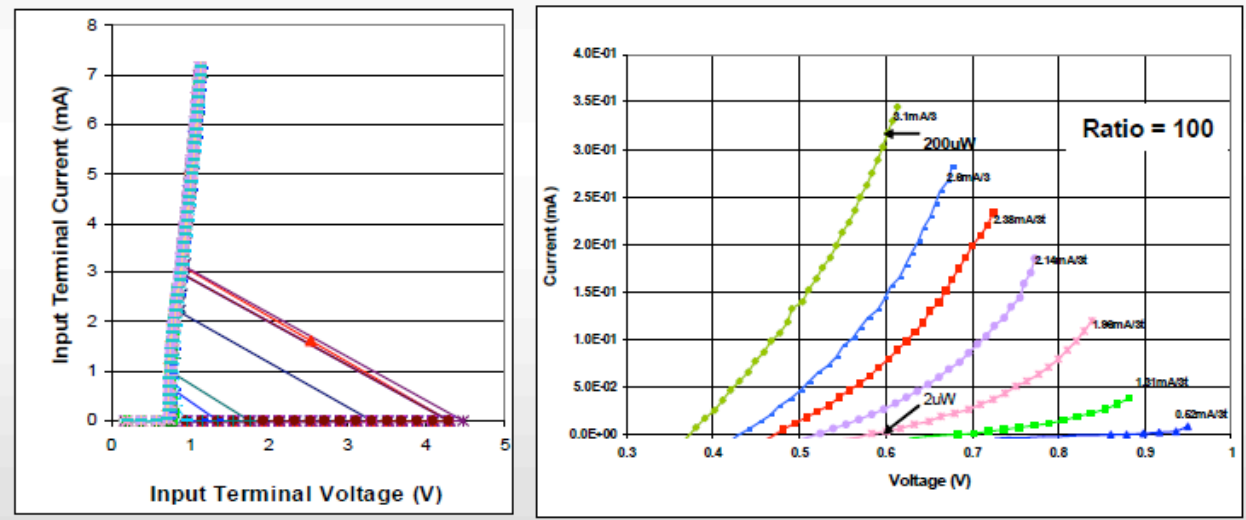

**Figure 4 – I(V) curves for a 3-terminal OTS device**

We are not going to repeat Chua's arguments here. Instead, we will discuss a version of the OTS device with three terminals, as shown in Figure 2. We will show that a three-terminal OTS device can both be used as an access device to an OMS bit instead of a transistor, and perform the inverse logical IMP operation. One can see from the I(V) curves in Figure 4 that we can modulate both the threshold voltage and the holding current of the device by applying bias on the third terminal.

In a recent paper, Intel has introduced the Ovonic Threshold Switch with phase change memory [1], shown in Figure 3. Now I will show that a 3-terminal Ovonic Threshold Switch can replace conventional transistors of all types with very definite advantages.

Figure 5 is an example of a three terminal OTS device operation in the so-called 'latching' mode:

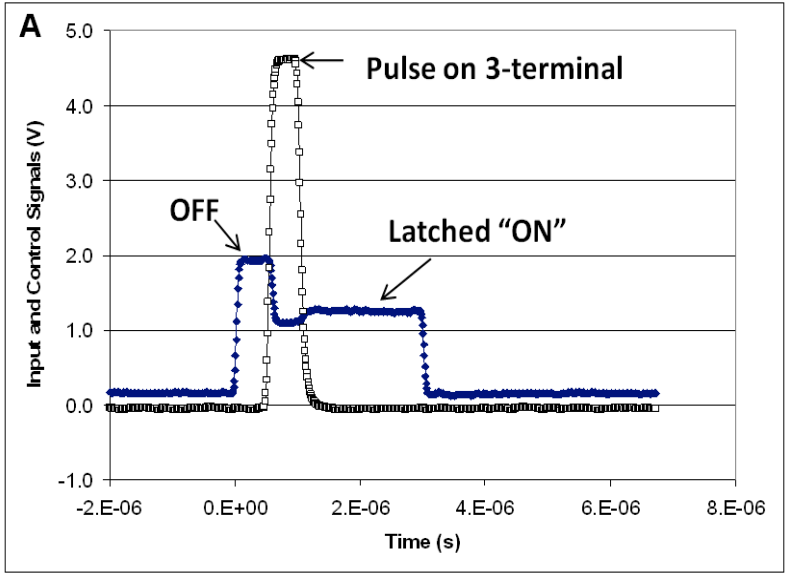

Figure 5 – 3-terminal OTS Latching Mode OTS operation

Figure 6 – Truth Table for a 3 Terminal device showing inverse IMP logic

From Figure 5 you can see that a 3-terminal threshold switch can be turned on by applying a short pulse on the intermediate terminal which latches the OTS in the on-state.

With the OTS in the on-state the memory device can be probed or programmed. It should be noted here, that because of that latching property, the 3-terminal threshold switch can be used as a volatile memory bit. Essentially the latching allows the OTS to be 'on' without any current through the third terminal, which also saves energy.

Because the voltage on the OTS can be switched from high to low (from sub-threshold to the holding voltage), the 3-terminal OTS can show the same IMP logic operation as the memory only essentially without storing it in a non-volatile fashion. Namely, the voltage top to bottom terminal can be high (1) or low (0) and the voltage on the 3$^{rd}$ terminal can be high (1) or low (0). Then, the output voltage top to bottom will be according to the truth table shown in Figure 6.

## 3/ The Ovonic Memory Switch

Much more uniqueness is available to the Ovonic Unified Memory/Cognitive Device (Figure 7).

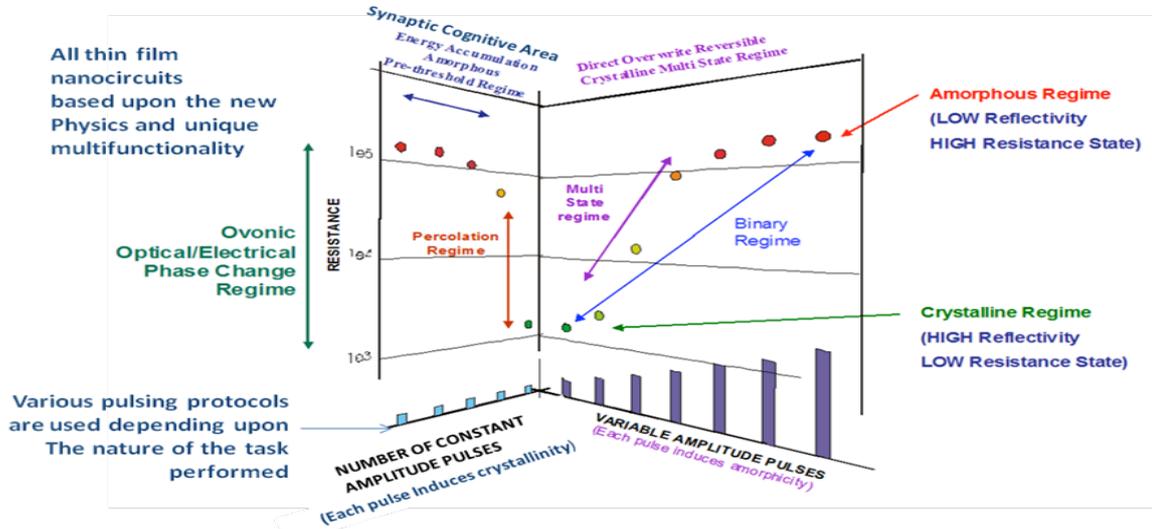

**Figure 7 – OMS device – cognitive and direct over-write mode**

As you can see from Figure 7, there are two regimes of operation of the OUM device – the one on the right is the direct overwrite regime used for non-volatile and multi-state memory. The one on the left is what we call 'the pre-threshold or cognitive regime'. If based on this figure we plot the memristance of the Ovonic Cognitive Device vs. the number of applied pulses, we get Figure 8 which shows that the Ovonic Cognitive Device acts as a memristor because the resistance of the device in the cognitive regime is a non-linear function of the total charge.    The non cognitive or direct over-write mode allows the resetting of the logical zero or 'false' state of the device, therefore the entire device acts as a neuron.

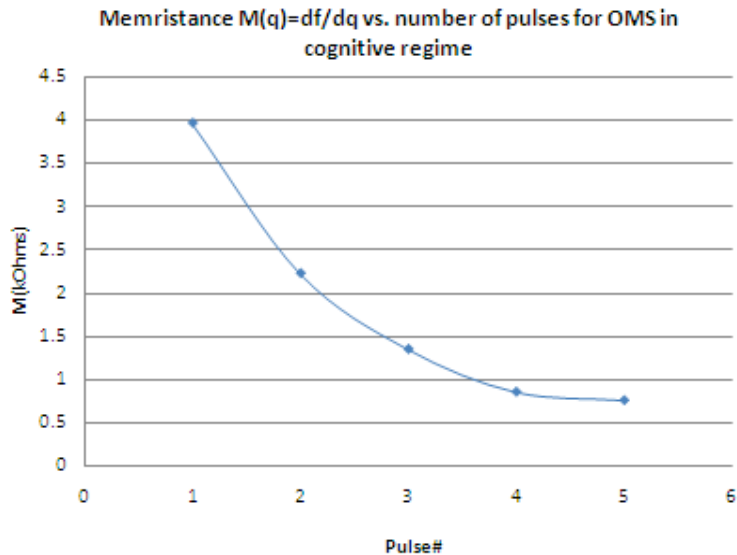

**Figure 8 – OMS in cognitive mode**

The device remembers the number of pulses that have been sent to it.   The way this information is stored is completely secure since the pre-threshold state cannot be forensically recognized.

There are a great number of patents in my name as well as my collaborator, Boil Pashmakov, related to three-terminal OTS and OMS devices. Please, see the references for the patent numbers at the end of this paper [10]. What is shown in these patents is that:

Ovonic devices can be used singly and together in circuits to provide a huge range of capabilities
- Can adapt some algorithms of quantum devices – ideal for factoring
- Addition, subtraction, division, multiplication
- Huge parallelism
- Weighted interconnections using multi-state phase change storage devices
- Search engine
  - Not just matching, but intelligent searches
  - Learns as you search
  - Associative capabilities

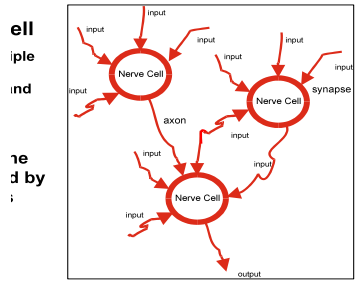

Ovonic Single and Multiple cells have the same properties as neurons and biological cells

Output fires when the threshold is reached by summing the inputs

The Ovonic Threshold switch does this instantly, continuously and reversibly

The Ovonic Cognitive Device does this through accumulation of input over time

*OVSHINSKY INNOVATION / OVONYX*

**Figure 9 – Cognitive Functionality**

More recently [2] Hewlett Packard came up with a claim for a memristor device that can be used as a non-volatile memory and as a logic device to perform logic by IMP ('material implication'). They show as in Figure 10, which was taken from their paper [2], that using 3 of their devices they can perform the logical IMP operation (compared to a *single* OMS device).

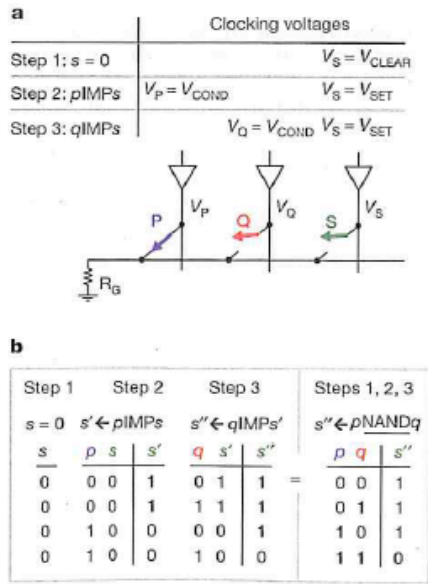

**Figure 10 – IMP logic using 3 memristors**

We have a patent for a three terminal OMS device [4], which, as shown in the truth table in Figure 11, can perform the same logic function as a single device. In this truth table, we have assigned logical '0'' to the 'reset' (amorphous or high resistance) state and logical '1' to the 'set' (crystalline or low resistance) state. We also input the logical variable 'p' in the top-to-bottom circuit and the logical variable 'q' in the 3t-to-top circuit. The result of the logical operation (q') is read as the output 3t-to-top.

| Input top-bottom (p) | Input 3t-top (q) | Output 3t-top(q') |
|---|---|---|
| Reset (0) | Reset (0) | Set (1) |
| Reset (0) | Set (1) | Set (1) |
| Set (1) | Reset (0) | Reset (0) |
| Set (1) | Set (1) | Set (1) |

q ' ⟵ pIMPq

**Figure 11 – Truth Table for a 3-terminal OMS device showing IMP logic operation**

Thus, we have shown that a three terminal OUM device can execute the logical operation "material implication" or IMP which combined with the logical false (or device reset), constitutes a ***computationally complete*** logic basis which means that all *Boolean* logic operations can be done using a ***single*** three terminal OUM device [2]. To see the neuro-physiological analogue of this device, please see: [5-7].

# 4. CONCLUSION

I have been writing and describing at the EPCOS meetings through the years that the Ovonic Phase Change Memory and Ovonic Threshold Switch and 3-terminal threshold switch are able to accomplish unusual fast switching and that the current density of the solid state plasma of the threshold switch is fifty times higher than conventional transistors. An incredible number that I have indicated and likened to a room temperature dynamic superconductor [8]. This is a discussion that I will reserve for the next meeting, but I just wanted to say again that the tremendous amount of lone pairs that are excited simultaneously with the Ohmic injection of carriers that fill the recombination centers created by the high electric field coupling to the lone pairs, creates conditions wherein there is probable boson generation and condensation to permit a superconducting possibility. Without going into detail one can take the number of excited carriers that can have in the plasma spin up and spin down character and sufficient density to permit a dynamic condensation process [8].

In ending, I hope that I have been able to show that there is unique electronic and phase change mechanisms where the Ovonic Memory has a threshold excitation action and wherein we can integrate in a single nano-device unique multifunctional circuits. Why is this important? I close by a very brief history: the transistor was invented and it exceeded expectations until it hit the numbers catastrophe where the wiring situation limited the addition of many more devices and this problem was solved by the invention of the integrated circuit. (By the way, because amorphous oxide was used which permitted fine lithography to be performed and one could have complex circuits on a single chip). What I am proposing in this paper is that now we have complex circuits in a single nano-device. For example, we have a unified memory and threshold action in which the unique mechanism will permit to perform logic and store data in the same device. It is important to mention and understand the vibronic nature of the polymeric lone pair chain structures of chalcogenide materials, wherein electronic transitions are made possible by the vibrational motion of the chains. These simultaneous vibrational and electronic transitions are important to understanding the mechanism of both the optical and the electrical Ovonic phase change memories [9].

**ACKNOWLEDGEMENTS**

I want to express my appreciation and thanks to Boil Pashmakov for his significant contributions to this work.

**Biography**

Stanford R. Ovshinsky has dedicated his life to creating an entirely new area of physics and materials science, providing innovation in atomic engineering of amorphous and disordered semiconductors.   In 1960, he and his late wife, Iris, founded the company Energy Conversion Devices, Inc. (ECD), to further develop and apply his inventions to the fields of information and energy creating a new field known as "Ovonics," attracting many scientists and technologists. The materials/devices developed by Ovshinsky resulted in unique switching devices, phase change memories, optoelectronic copying, and flat-panel liquid crystal displays.   In the information field, he has focused on using the Ovonic phase change memories and Ovonic threshold switches for 3-terminal devices and cognitive computing.   Mr. Ovshinsky holds over 400 U.S. patents, some of which led to remarkable new approaches to the uses of solar power and hydrogen fueled vehicles. His battery technology enabled electric and hybrid vehicles. His patents for a system that allows photovoltaic solar panels to be manufactured in long continuous rolls provided a revolutionary leap for solar energy.   He has formed a new company named Ovshinsky Solar LLC in order to accelerate his work in energy, leading to basic solutions for pollution, climate change gases, and the need for oil.   His objective now is to make possible photovoltaics at a lower cost than burning fossil fuel.   Mr. Ovshinsky has authored over 300 scientific papers, and has received global recognition for his discoveries, including the Diesel Gold Medal for Invention, presented by the Deutscher Erfinerverband (German Inventors Association), the 2005 Innovation Award for Energy and the Environment by the British publication *The Economist*, the 2008 Engineering Society of Detroit Lifetime Achievement Award.   He was named "Hero for the Planet" by *Time* magazine and with Iris named Heroes of Chemistry by the American Chemical Society.   He has numerous honorary degrees, most recently receiving an Honorary Doctor of Science from the University of Michigan (2010), and is a fellow of the American Physical Society and the American Association for the Advancement of Science.